\documentclass[english,american]{article}
\usepackage[T1]{fontenc}
\usepackage[latin9]{inputenc}
\usepackage{geometry}
\usepackage{float}
\usepackage{graphicx}
\usepackage{setspace}
\usepackage{babel}
\usepackage{epsfig}

\makeatletter
\@ifundefined{date}{}{\date{}}
\makeatother

\begin{document}

\title{The Topology of Inter-industry Relations from the Portuguese National
Accounts}
\author{Tanya Ara\'{u}jo$^{1}$* and Rui Faustino\\
ISEG (School of Economics and Management), University of Lisbon\\
Miguel Lupi 20, 1248-079 Lisbon\\
$^{1}$ UECE - Research Unit on Complexity and Economics}
\date{}
\maketitle

\begin{abstract}
In last years, the Portuguese economy has gone through a severe adjustment
process, affecting almost all industrial sectors, the building blocks of
economic structures. Research on economic structural changes has made use of
input/output tables to define networks of industrial relations. Here, these
networks are induced from output tables of the Portuguese national
accounting system, being each inter-industry relation defined by the output
made by any two industries for the products that they both produce. The
topological analysis of these networks allows to uncover a particular
structure that comes out during the Portuguese adjustment program. The
evolution of the industrial networks shows an important structural change in
2011-2014, confirming the usefulness of inducting similarity networks from
output tables and the consequent promising power of the graph formulation
for the analysis of inter-industry relations.

\begin{description}
\item[Keywords:] Industry/product Output Table, Network Analysis, Minimal
Spanning Trees, Industrial Clusters, National Accounting Systems, Proximity
Networks
\end{description}
\end{abstract}

{\small * Corresponding author (tanya@iseg.ulisboa.pt)}

{\small Financial support from national funds by FCT (Funda\c{c}\~{a}o para
a Ci\^{e}ncia e a Tecnologia). This article is part of the Strategic
Project: UID/ECO/00436/2013.}

\section{Introduction}

As most complex systems, economic structures may be described in many
different ways. The simplest descriptions are usually built on top-down
decompositions, where aggregation of economic concerns can be either driven
by institutional sectors: households, firms and government or by economic
outputs as in the three approaches of GDP measurement: the production,
expenditure and income approaches.

The economic outputs used in the calculation of GDP by the production
approach are collected, validated and reported by national statistical
systems. In so doing, these systems organize production by products (goods
and services) which are produced by industries. Because each industry is
able to produce several goods and since a given good can be the output of
various industries, the top-down decomposing chain (sector $\rightarrow $
industry $\rightarrow $ product) may be conveniently replaced by a bottom-up
description, where economic structures are built on inter-industry
relations. In this setting, it is possible to define each inter-industry
relation as the output made by any two industries for the products that they
both produce. In so doing, the strength of an inter-industry relation -
between any two industries - depends on the number of common products that
they share.

The adoption of a bottom-up perspective and the availability of year-based
data allows for analyzing the time evolution of inter-industry relations
from a network approach. Moreover, such a relational setting provides the
basis for evaluating the degree of specialization of the economy from the
distribution of its industrial production.

Network approaches are the natural setting for representing and analysing
relational linkages. There has been an increasing interest in applying
network approaches to economic problems, from the reconstruction of
artificial financial markets to the analysis of international trade, there
is a great number of successful and inspiring applications (\cite{Hid07}-%
\cite{Lee11}).The first step in the adoption of a network approach concerns
the definition of the network nodes and links. As there are many ways to
relate the elementary units of a system, the choices may depend strongly on
the questions that a network analysis aims to address \cite{Ara16a}.

The main objective of this paper is to investigate the extent to which the
Portuguese economic performance from 2000 to 2014 had some bearing on the
Portuguese inter-industry relations. Given the recent process of strong
economic adjustment suffered by the Portuguese economy from 2011 to 2014, we
focus on the impact of this process on the Portuguese economic structure
which is herein represented by bipartite networks of inter-industry
relations.

In many economic networks\ - and specially in those induced from empirical
data - the adoption of a network representation intuitively emerge. It
happens because these systems are characterized by a low abstraction level,
being the network representation the most obvious solution, as in the case
of air-traffic, power-grid, and trade networks.

It also happens with the specific field of input/output (I/O) tables, an
important part of the national accounting systems. Because I/O tables are
quite similar to adjacency matrices there has been an increased interest in
applying network theory to represent money flows between industrial sectors
\cite{Blo11}-\cite{Aro03}.

In the pioneering work of Slater \cite{Sla77}, 75 industries are clustered
according to the USA inter-industry flow table of 1967. Later, Schnabl \cite%
{Sch94} applies Minimal Flow Analysis to induce networks from the German I/O
tables reporting data in between 1978 and 1988. There, centrality measures
allow for classifying industries into three different sectors (source, sink
and center). More recently, Bl\"{o}chl and co-authors \cite{Blo11},\ using
I/O tables of 37 OECD countries, induce networks of industries to which
measures of random walk centrality and counting betweenness are applied.
Their results have shown the suitability of those measures in the
identification of groups of countries according to their development status.

The present study also falls into the broad category of data-driven
investigation on industrial relations using a network approach.
Nevertheless, we follow a different perspective. Instead of considering I/O
tables, we take the output of each industry distributed by the set of
products that this industry produces. Inter-industry relations are then
defined by the production of common products. Industries are linked whenever
they share at least one mutual product, being the strength of each
inter-industry link defined by the output made by the involved industries
for the products that they both produce. In so doing, the intensity of a
link between any two industries depends on the number of mutual products
weighted by their relative (output) values in each linked industry.

The networks we work with are proximity (and bipartite) networks. In
proximity networks, the links are defined from shared features, correlation
coefficients or other well-defined similarity measures. Like in many other
economic networks, the elementary units do not have to be explicitly linked
by any concrete relation existing in the real world except for a
well-defined measure of distance in between them. Although the induction of
proximity networks is less intuitive than those obtained from the
air-traffic, power-grid or I/O table examples, they provide useful
analytical settings, being found in a multitude of applications. Examples of
proximity networks in Economics can be found in references \cite{Ara07} ,%
\cite{Ara16b} and \cite{Spe12}. A detailed discussion on proximity networks
is presented in reference \cite{Ara16a}.

The paper is organized as follows. Section two presents the data we work
with. Section three is targeted at presenting the methodological aspects and
a brief discussion on the first results. Section four discuss the evolution
of the Portuguese industrial networks, focussing on the topological analysis
of different time periods . Finally, Section five presents the concluding
remarks.

\section{The data}

Our data source is the year-based Industry/Product output tables ($OT^{t}$ $%
t=2000,2001,...,2014$) compiled by the Portuguese national accounting agency
(INE \cite{INE16}). The output tables consist of data on production values
(at market prices) organized by industry and related products. In this
context, industries (I) refer to firms and other business, and the products
(P) refer to goods and services. Moreover, while the classification of a
given business into a specific industrial category (I) is determined by its
economic activity classification (NACE code) by taking into account the main
activity of the firm; products (P) are classified by activity according to a
statistical coding (CPA code).

There are international standards for the industry (and product)
classification sets provided by the United Nations (UN) and adopted by most
of countries. The UN System of National Accounts (SNA) provides the basis
for uniformity among the various data sets, while the OECD industry
classifying into 10, 21 and 38 industries makes the international
comparisons possible \cite{Kel08}.

Table 1 shows a snapshot of the classifying list of industries and products
according to NACE and CPA codes at the resolution of 38 industries and 38
products (I38 \& P38). Table A (in the appendix) provides a complete
presentation of the industries and products at the same resolution.

Because the classification of each industry is determined by its main
economic activity, which in turn is determined by the type of goods or
services it produces, the classification of industries happens to be
identical to the classification of products, as the last two columns of
Table 1 show. Moreover, except for the public administration industry (O),
the production of every industry relies mainly on a specific product whose
code is naturally inherited from the industry itself. Even though,
industries do produce several other products, besides the main (self-coded)
products, each of which happens to be the output of various industries.

\begin{center}
\begin{tabular}{|l|l|l|l|}
\hline
{\small NACE and } & {\small Description} & {\small I38} & {\small P38} \\
\cline{2-4}
{\small CPA code} &  &  &  \\ \hline
{\small 01-03} & {\small Agriculture, forestry and fishing } & {\small A} &
{\small A} \\ \hline
{\small 05-09} & {\small Mining and quarrying} & {\small B} & {\small B} \\
\hline
{\small 10-12} & {\small Food products, beverages and tobacco products} &
{\small CA} & {\small CA} \\ \hline
{\small 13-15} & {\small Textiles, wearing apparel, leather and related
products} & {\small CB} & {\small CB} \\ \hline
{\small 16-18} & {\small Wood and paper products, and printing} & {\small CC}
& {\small CC} \\ \hline
{\small ...} & {\small ...} & {\small ...} & {\small ...} \\ \hline
{\small 41-43} & {\small Construction} & {\small F} & {\small F} \\ \hline
{\small 45-47} & {\small Wholesale and retail trade } & {\small G} & {\small %
G} \\ \hline
{\small ...} & {\small ...} & {\small ...} & {\small ...} \\ \hline
{\small 64-66} & {\small Financial and Insurance} & {\small K} & {\small K}
\\ \hline
{\small 68} & {\small Real estate} & {\small L} & {\small L} \\ \hline
{\small ...} & {\small ...} & {\small ...} & {\small ...} \\ \hline
{\small 97} & {\small Households as employers of domestic personnel} &
{\small T} & {\small T} \\ \hline
{\small 99} & {\small Activities of extraterritorial organizations and bodies%
} & {\small U} & {\small U} \\ \hline
\end{tabular}

{\small Table 1: A snapshot of the list of industries and products according
to NACE and CPA codes at the resolution I38 \& P38.}
\end{center}

Our approach is applied to a data set comprising 15 output tables ($%
OT_{38}^{t}$ $t=2000,2001,...,2014$) at the resolution of 38 industries and
38 products (I38 \& P38) reported by INE. Table 2 shows part of the
Industry/Product output table of 2000 ($OT_{38}^{2000}$), a framework where
each industry is associated with the set of products it produces. Each cell
in $OT_{38}^{2000}$\ represents the value $v^{^{2000}}(i,p)$ of the output $%
p $ produced by industry $i$ in 2000.

\begin{center}
\begin{tabular}{|l|l|l|l|l|l|l|l|}
\hline
& I$_{38}$/P$_{38}$ & P$_{1}$ & P$_{2}$ & ... & P$_{36}$ & P$_{37}$ & P$%
_{38} $ \\ \hline
&  & A & B & ... & S & T & U \\ \hline
I$_{1}$ & A & {\small \ 6375238 } & {\small 96} & {\small ...} &  &  &
{\small -} \\ \hline
I$_{2}$ & B & {\small 0} & {\small \ 837319 } & {\small ...} &  &  & {\small %
-} \\ \hline
... & ... & {\small ...} & {\small ...} & {\small ...} & {\small ...} &
{\small 0} & {\small -} \\ \hline
I$_{36}$ & S & {\small 0} & {\small 0} & {\small ...} & {\small \ 1513029 }
& {\small 0} & {\small -} \\ \hline
I$_{37}$ & T & {\small 0} & {\small 0} & {\small 0} & {\small 0} & {\small \
494448 } & {\small -} \\ \hline
I$_{38}$ & U & {\small -} & {\small -} & {\small -} & {\small -} & {\small -}
& {\small -} \\ \hline
\end{tabular}

{\small Table 2: Part of the industry/product output table at the resolution
I38 \& P38 (values in 10}$^{{\small 3}}${\small \ euros).}
\end{center}

Since our main focus relies on the analysis of the most recent years, Table
3 shows the number of products ($\#p$) by industry in 2010 and 2014. Because
some products are not produced at a significant level, we added columns (\#$%
p ${\small {>}10}$^{{\small 6}}$) with the number of products
whose output values are above one million euros. When considering the
significant levels of production, the vast majority of industries produced
between 6 and 15 products. In opposing situations are the industries G (%
\textit{Wholesale trade}) and T (\textit{Goods and services producing
activities of households for own use}) presented at the last row of Table 3.
The former produces just one product while the latter, produces more than 30
products, both in 2010 and 2014.

\begin{center}
\begin{tabular}{lccccccccc}
{\small Industry} & 2010 &  & 2014 &  & {\small Industry} & 2010 &  & 2014 &
\\ \hline
& $\#p$ & \#$p${\small {>}10}$^{{\small 6}}$ & $\#p$ & \#$p$%
{\small {>}10}$^{{\small 6}}$ &  & $\#p$ & \#$p${\small
{>}10}$^{{\small 6}}$ & $\#p$ & \#$p${\small {>}10}$%
^{6}$ \\
\textbf{A} & {\small 16} & {\small 10} & {\small 15} & {\small 10} & \textbf{%
H} & {\small 15} & {\small 14} & {\small 15} & {\small 12} \\
\textbf{B} & {\small 11} & {\small 7} & {\small 10} & {\small 7} & \textbf{I}
& {\small 12} & {\small 10} & {\small 13} & {\small 9} \\
\textbf{CA} & {\small 12} & {\small 9} & {\small 12} & {\small 7} & \textbf{%
JA} & {\small 16} & {\small 10} & {\small 15} & {\small 8} \\
\textbf{CB} & {\small 20} & {\small 14} & {\small 17} & {\small 12} &
\textbf{JB} & {\small 12} & {\small 10} & {\small 12} & {\small 10} \\
\textbf{CC} & {\small 19} & {\small 16} & {\small 19} & {\small 13} &
\textbf{JC} & {\small 11} & {\small 7} & {\small 12} & {\small 7} \\
\textbf{CD} & {\small 7} & {\small 6} & {\small 6} & {\small 6} & \textbf{K}
& {\small 7} & {\small 7} & {\small 6} & {\small 6} \\
\textbf{CE} & {\small 16} & {\small 13} & {\small 15} & {\small 12} &
\textbf{L} & {\small 10} & {\small 8} & {\small 8} & {\small 6} \\
\textbf{CF} & {\small 11} & {\small 6} & {\small 10} & {\small 6} & \textbf{%
MA} & {\small 14} & {\small 10} & {\small 12} & {\small 9} \\
\textbf{CG} & {\small 20} & {\small 15} & {\small 20} & {\small 16} &
\textbf{MB} & {\small 14} & {\small 5} & {\small 15} & {\small 7} \\
\textbf{CH} & {\small 20} & {\small 14} & {\small 20} & {\small 14} &
\textbf{MC} & {\small 13} & {\small 6} & {\small 12} & {\small 8} \\
\textbf{CI} & {\small 16} & {\small 12} & {\small 15} & {\small 12} &
\textbf{N} & {\small 10} & {\small 6} & {\small 10} & {\small 6} \\
\textbf{CJ} & {\small 17} & {\small 12} & {\small 16} & {\small 13} &
\textbf{O} & {\small 28} & {\small 20} & {\small 27} & {\small 21} \\
\textbf{CK} & {\small 17} & {\small 12} & {\small 16} & {\small 12} &
\textbf{P} & {\small 24} & {\small 13} & {\small 21} & {\small 13} \\
\textbf{CL} & {\small 17} & {\small 14} & {\small 16} & {\small 13} &
\textbf{QA} & {\small 19} & {\small 11} & {\small 17} & {\small 12} \\
\textbf{CM} & {\small 20} & {\small 18} & {\small 20} & {\small 17} &
\textbf{QB} & {\small 24} & {\small 9} & {\small 19} & {\small 9} \\
\textbf{D} & {\small 10} & {\small 6} & {\small 10} & {\small 6} & \textbf{R}
& {\small 18} & {\small 10} & {\small 17} & {\small 10} \\
\textbf{E} & {\small 16} & {\small 11} & {\small 15} & {\small 11} & \textbf{%
S} & {\small 11} & {\small 7} & {\small 11} & {\small 7} \\
\textbf{F} & {\small 15} & {\small 15} & {\small 18} & {\small 15} & \textbf{%
T} & {\small 1} & {\small 1} & {\small 1} & {\small 1} \\
\textbf{G} & {\small 32} & {\small 32} & {\small 32} & {\small 32} & \textbf{%
U} & {\small 0} & {\small 0} & {\small 0} & {\small 0} \\ \hline
\end{tabular}

{\small Table 3: Number of products by industry and number of products
produced at a significant level, at the resolution  I38 \& P38.}
\end{center}

Figure 1 shows the distribution of the number of products by industry in
2010 and 2014, together with the information provided in Table 3, those
distributions suggest that:

\begin{enumerate}
\item there is a slight increase in diversification of production from 2010
to 2014

\item the number of products produced by industries G (\textit{Wholesale
trade}) and CC, CG, CF, CH and CM (manufacturing industries) remains almost
unchanged (Figure 1(a))

\item the number of products produced by industries F (\textit{Construction}%
) increases significantly from 2010 to 2014 but the number of those produced
at a significant level remain unchanged

\item the strongest difference between the number of products and those
produced at a significant level relies on the industries O (\textit{Public
administration}), P (\textit{Education}), QA (\textit{Human health}) and QB (%
\textit{Social work})

\item the highest decreases from 2010 to 2014 in the number of products
produced at a significant level (Figure 1(b)) occurs in industries CB (%
\textit{Textiles}) and L (\textit{Real state})

\item among the few industries that display an increase in the number of
products produced at a significant level are MB (\textit{R\&D}) and MC (%
\textit{Other professional, scientific and technical services})
\end{enumerate}

\begin{figure}[tbh]
\begin{center}
\psfig{figure=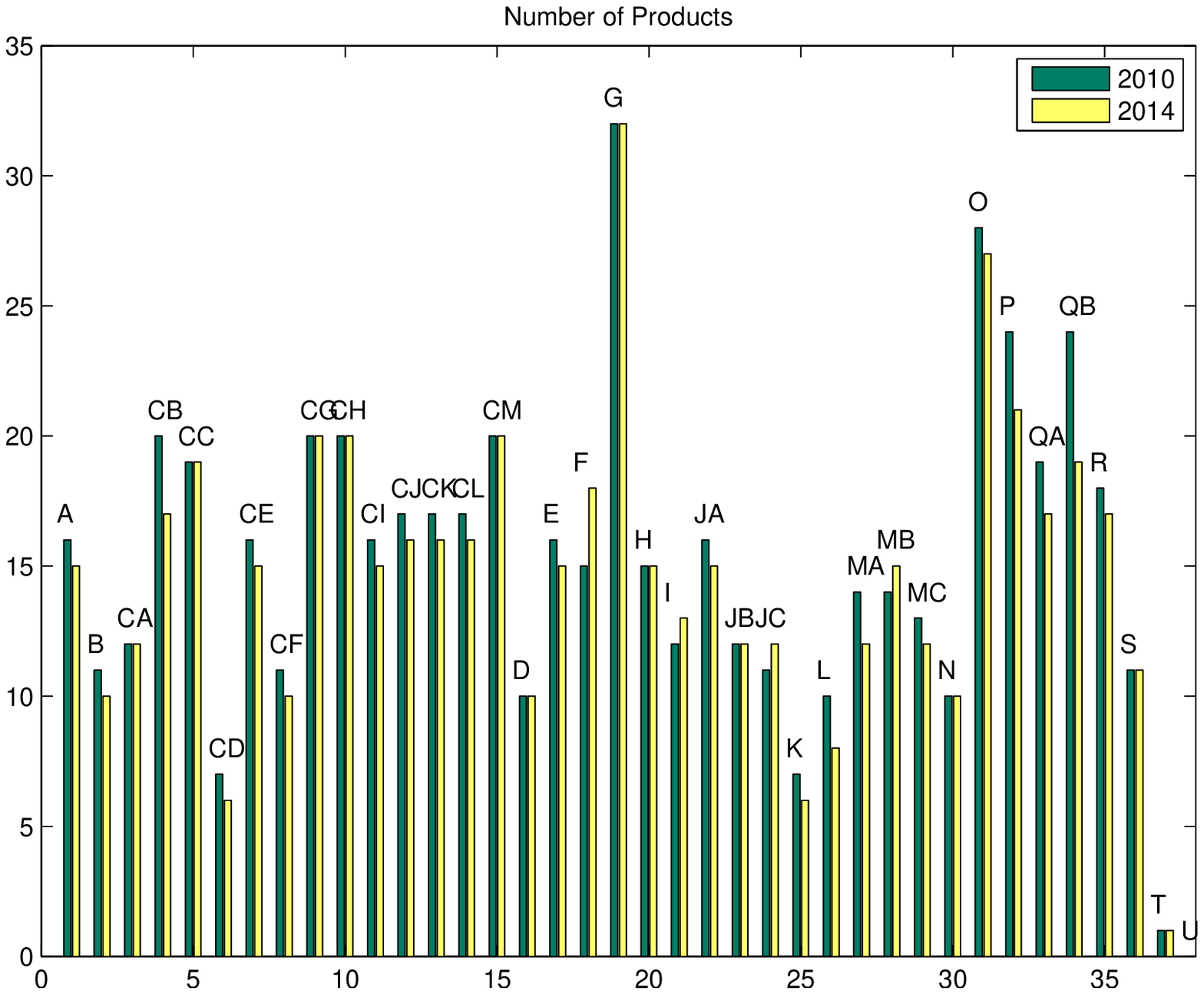,width=8truecm}
\psfig{figure=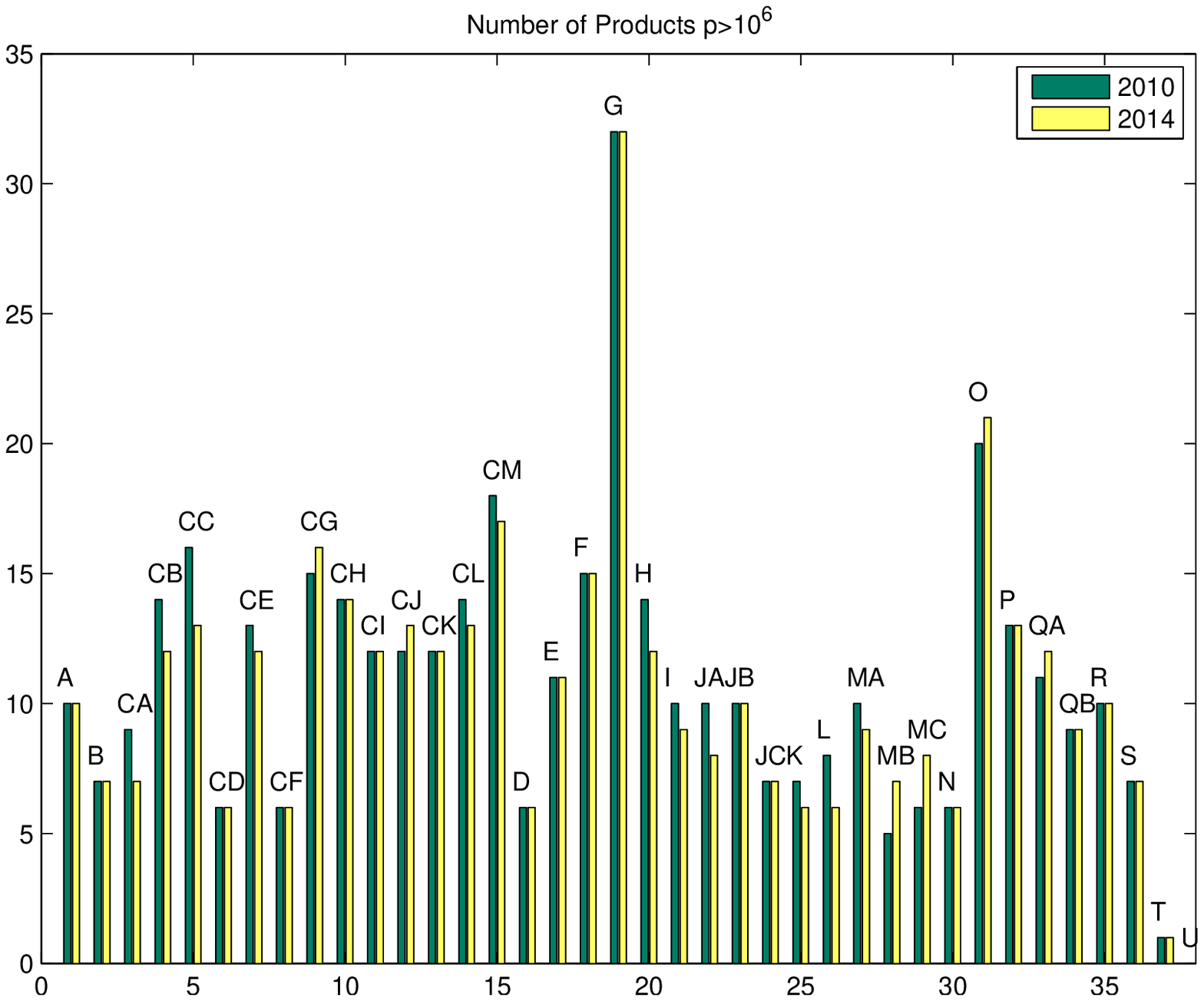,width=8truecm}
\caption{a: The number of products by industry at the
resolution of 38 industries and 38 products in 2010 and 2014 and b: The number of products produced at a significant level, at the resolution of
38 industries and 38 products in 2010 and 2014.}
\end{center}
\end{figure}

\section{Networks}

As earlier mentioned, in defining any network, there are many different
design decisions to be taken. The choice of a given set of nodes and the
definition of the links between them is only one out of several other ways
to look at a given system. Here we define year-based bipartite networks
where similarities between industries are used to set the existence of every
link in each network. Being weighted graphs, the weight of each link is
proportional to the intensity of the similarity between the linked pair of
industries, relative to the overall output value of each involved industry.

Because those bipartite networks have a large number of links, we compute
their corresponding Minimal Spanning Trees (MST). In fact, it happens that
when networks are induced from similarity measures, the issue of deriving a
sparse network from a dense or even a complete one becomes meaningful. The
less arbitrary choices (or the most endogenously based ones) usually relies
on the construction of a MST. In so doing, we ensure the connectivity is
preserved (the resulting network is necessarily connected) while moving from
a dense network to a sparse one. Moreover, we are able to emphasize the main
topological patterns that emerge from the network representations. Beyond
the MST analysis, we evaluate and discuss \textit{the amount of redundancy}
in the network links and the values of the \textit{residuality }coefficient
computed for each year-based network of industries. These coefficients -
redundancy and residuality - help to characterize the main differences in
the behavior of those networks in the latest years.

\subsection{Bipartite Graphs}

A bipartite network $N$ consists of two partitions of nodes $I$ and $P$,
such that edges connect nodes from different partitions, but never those in
the same partition. A projection of such a bipartite network onto $I$ is a
network consisting of the nodes in $I$ such that two nodes $i$ and $i\prime $
are connected, if and only if there exist a node $p\in P$ such that $(i,p)$
and $(i\prime ,p)$ are edges in the corresponding bipartite network ($N$).

Here, the two partitions of nodes $I$ and $P$ are the set of industries and
the set of products, respectively, both at the resolution of 36 elements%
\footnote{%
Our networks have 36 nodes instead 38, because industries T and U were
excluded since U has no data and the T produces only one product, remaining
therefore without any inter-industry relation.}, as presented in Table A1
(and partially displayed in Table 2). The links between any two industries $%
(i,i\prime )$ in \ the network $N$ are defined by the existence of products (%
$p\in P$) such that $(i,p)\in $ $N$ and $(i\prime ,p)\in $ $N.$

In the following, we explore bipartite networks and their corresponding
projections $N_{36}^{t}(i,p)$ where $t=2000,2001,...,2014$ with $i$ $\in $
\{I$_{1}$,I$_{2}$ ,..., I$_{36}$\} and $p\in $ \{P$_{1}$,P$_{2}$ ,..., P$%
_{36}$\}.

\subsection{Networks of Industries}

Given that each industry can produce many products and that each product can
be produced by several industries, from each output table, the values $%
v^{t}(i,p)$ $t=\{2000,2001,...,2014\}$ with $i$ $\in $ \{I$_{1}$,I$_{2}$
,..., I$_{36}$\} and $p\in $ \{P$_{1}$,P$_{2}$ ,..., P$_{36}$\} relating
industries to products are taken and the proximity networks $N_{36}^{t}$ are
induced. There, nodes are industries (I$_{i}$ $i$ $\in $ \{I$_{1}$,I$_{2}$
,..., I$_{36}$\} and links $n_{(a,b)}$ are defined by:

\begin{equation}
n_{(a,b)}=\sum_{k=1}^{36}p_{ak}p_{bk}
\end{equation}

where the $p_{ik}$ and $p_{jk}$ are the normalized values of the outputs of
industries $i$ and $j$ for the product $k$, respectively.

The values $v(a,k)$ of the product $k$ produced by industry $a$ are
normalized by industry, summing up the output values of all the products ($%
k) $ that industry $a$ produces:

\begin{equation}
p_{ak}=\frac{v(a,k)}{\sum \limits_{k=1}^{k=36}v(a,k)}10^{3}
\end{equation}

Therefore, the higher is the value of the mutual production of two
industries (nodes $I_{a}$ and $I_{b}$), the greater is the strength of the
connection ($n_{(a,b)}$) between industries $a$ and $b$.

As an example, and taking from the output table $OT_{38}^{2014}$ the
normalized values (Eq.2) of each mutual product produced by industries QA (%
\textit{Human health}) and QB (\textit{Social work}) in 2014 yields\ (Eq.1) $%
n_{QA,QB}=0.84$ while, for instance, $n_{CI,CJ}=94.65$, since the normalized
values (Eq.2) of each mutual product produced by industries CI (\textit{%
Manufacture of computer, electronic and optical products}) and CJ (\textit{%
Manufacture of electrical equipment}) in 2014 gives (Eq.1) $94.65$ . Not
surprisingly, the link between industries CI (\textit{Manufacture of
computer, electronic and optical products}) and CJ (\textit{Manufacture of
electrical equipment}) is around 100 times stronger than the connection
between industries QA (\textit{Human health activities}) and QB (\textit{%
Social work activities}), as the second graph in Figure 2 shows.

A multitude of different connection strengths can be observed in the
networks N$_{36}^{2000}$ and N$_{36}^{2014}$ presented in Figure 2. They are
weighted networks where the width of the links is proportional to the
strength of the connection between the involved nodes, and the size of the
nodes is proportional to their Gross Value Added (GVA)\footnote{%
The Gross Value Added (GVA) = Production - Intermediate Consumption
represents the weight of the industry in the national GDP. Some industries
with significant production values may have a reduced value added, as for
instance, industry CD (\textit{Coke and refined petroleum products}).}.

The first graph in Figure 2 (Figure 2(a)) shows that industries G (\textit{%
Wholesale trade}), O (Public \textit{administration}), F (\textit{%
Construction}) and L (\textit{Real state}) are those with the highest GVA
(the largest nodes) in the entire graph. The network at the right side of
Figure 2 (Figure 2(b)) shows that industries MB (\textit{R\&D}) \ and N\ (%
\textit{Education}) display increased sizes (larger GVA) when compared with
their situation in 2000. Moreover,\ the industries F (\textit{Construction})
and K (\textit{Financial services}) are among those with the highest
decrease of GVA in 2014. It happens together with the significant decrease
in the number of products of industry F in 2014 when compared with 2000.
Figure 2 also shows that the link between industries CI (\textit{Manufacture
of computer, electronic and optical products}) and CJ (\textit{Manufacture
of electrical equipment}) is largely reinforced in 2014, while the
connection between CF (\textit{Pharmaceuticals}) and G (\textit{Wholesale
trade}) loses weight when compared with 2000.

The overloaded graphs presented in Figure 2 do not favor the observation of
any particular pattern besides those associated with the size of the nodes
and links in each of those years.

\begin{figure}[tbh]
\begin{center}
\psfig{figure=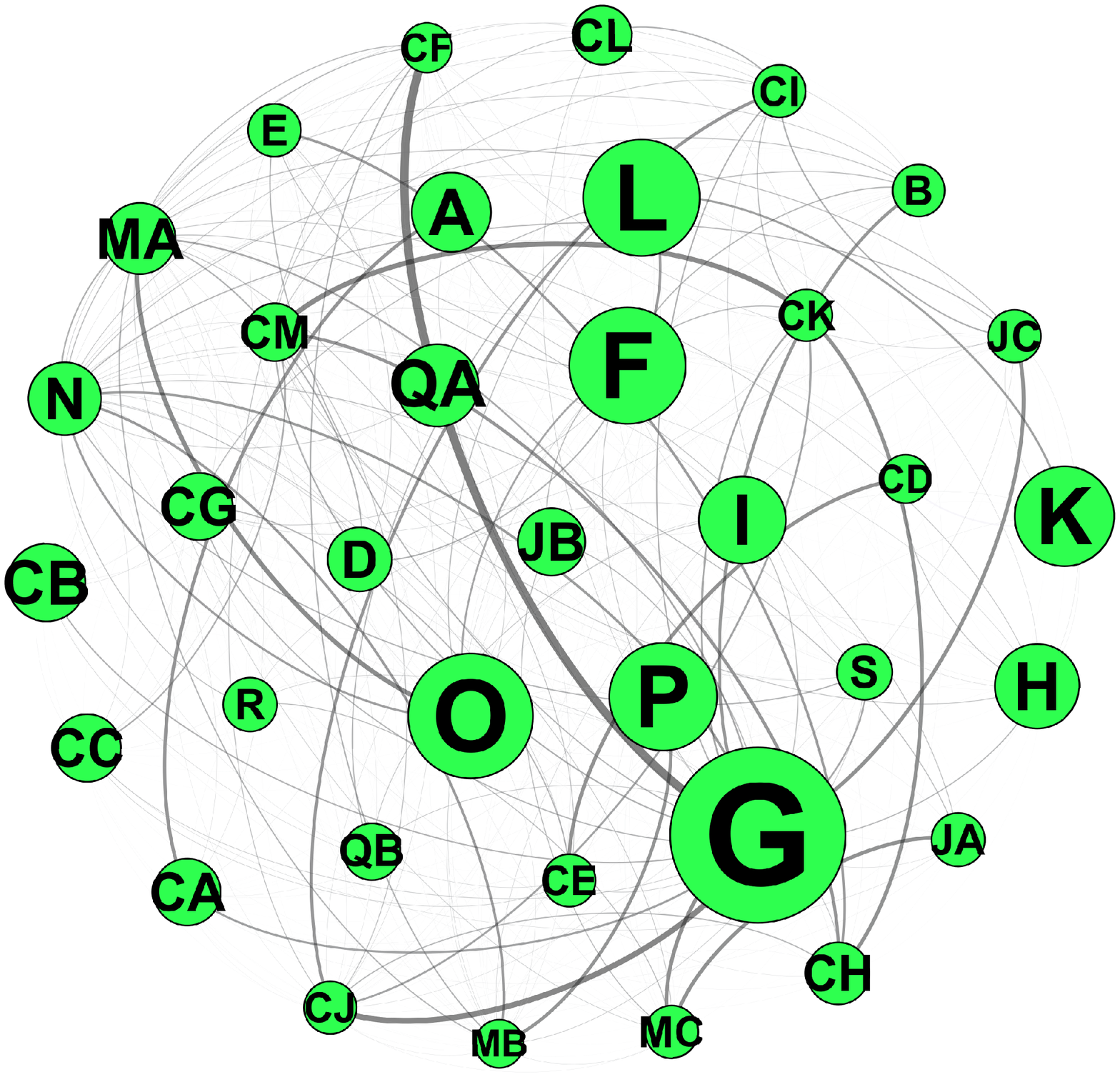,width=7truecm} \psfig{figure=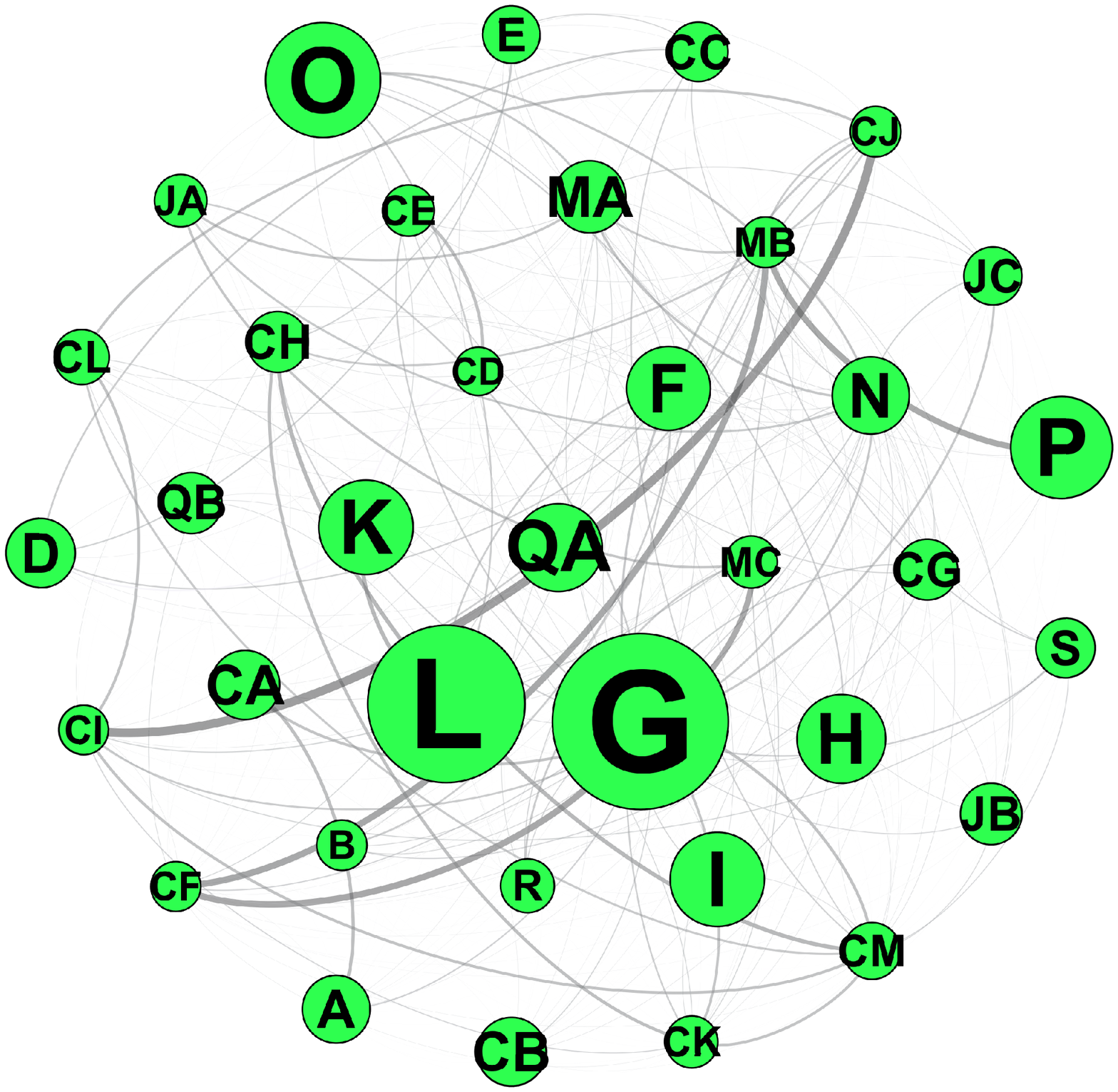,width=7truecm}
\caption{The networks of industries $N_{36}^{2000}$ and $N_{36}^{2014}$.}
\end{center}
\end{figure}

Undoubtedly, the networks above presented are not very informative about any
emerging structure in the network of industries. As earlier mentioned, when
networks are induced from similarity measures, in deriving a sparse network
from a dense one, the less arbitrary choices (or the most endogenously based
ones) usually rely on filtering the complete network with the threshold
distance value used in the last step of the hierarchical clustering process
of the construction of a Minimum Spanning Tree (MST). In so doing, we ensure
the connectivity is preserved since the resulting network is necessarily
connected.

\subsection{Minimum Spanning Trees}

A MST of a connected and weighted graph can be constructed by taking its
nodes and links and applying the \textit{nearest neighbor} method. The first
step in this direction is the computation of the distances $d_{ij}$ between
each pair of nodes $i$ and $j$ as the inverse of the weight of the link
between them:

\begin{equation}
d_{ij}=\frac{1}{n_{ij}}
\end{equation}

From the distance matrix $D_{36}^{t}$ a hierarchical clustering is then
performed. Initially $36$ clusters corresponding to the 36 industries are
considered. Then, at each step, two clusters $c_{i}$ and $c_{j}$ are clumped
into a single cluster if

\begin{center}
$d\{c_{i},c_{j}\}=\min \{d\{c_{i},c_{j}\}\}$
\end{center}

with the distance between clusters being defined by

\begin{center}
$d\{c_{i},c_{j}\}=\min \{d_{pq}\}$ with $p\in c_{i}$ and $q\in c_{j}$
\end{center}

This process is continued until there is a single cluster. In a connected
graph with $n$ nodes, the MST is a tree of $n-1$ edges that minimizes the
sum of the edge distances. In a network with $36$ nodes, the hierarchical
clustering process takes $35$ steps to be completed, and uses, at each step,
a particular distance $d_{i,j}$ $\in $ $D_{36}^{t}$ to clump two clusters
into a single one.

Let $C=\{d_{q}\}$, $q=1,...,m$, be the set of distances $d_{ij}\in
D_{36}^{t} $ used at each step of the clustering and the threshold distance $%
L_{36}^{t}=\max \{d_{q}\}$. After the last step, we are able to define a
representation of $D_{36}^{t}$ with sparseness replacing high-connectivity
in a suitable way. In the next section, we analyze the evolution of the
threshold distance $L_{36}^{t}$ $\ t=2000,2001,...,2014$ with particular
attention to the behavior of $L_{36}^{t}$ after 2010.

\section{Results}

Figure 3 shows the MST of the networks N$_{36}^{2000}$, N$_{36}^{2005}$
where the size of the nodes remain defined as in Figure 2, being
proportional to the yearly amount of GVA of each industry. Here, nodes are
colored according to their partition clusters, computed by a community
detection algorithm \cite{Blo08} available at Gephi, an open source software
for exploring and manipulating networks (https://gephi.org).

Four partition clusters are defined, they are:

\begin{enumerate}
\item \textsc{Trade} (\emph{green}): is the larger cluster, comprising
industries A (\textit{Agriculture and fishing)}, CA (\textit{Food products
and beverages)}, and CB (\textit{Textiles)}.

\item \textsc{Construction} (\emph{red}): is the most adversely affected
cluster after 2005. Besides industry F (\textit{Construction}), it comprises
L (\textit{Real estate}) and K (\textit{Financial services}).

\item \textsc{Manufacture }\emph{(blue)}: is the most stable cluster,
comprising CH (\textit{Basic metals}), CK (\textit{Machinery}) and CM (%
\textit{Furniture}).

\item \textsc{Arts} \emph{(yellow)}: is the less stable cluster. It is made
of \ industries R (\textit{Arts}), JA (\textit{Publishing}) and MC (\textit{%
Scientific and technical services}).
\end{enumerate}

In the networks presented in Figures 3 and 4, white nodes represent those
whose inclusion into a specific cluster is largely unstable over the 15
years period 2000-2014. They are made by industries CG (\textit{Rubber and
plastics}), CE (\textit{Manufacture of chemical services}) and CD (\textit{%
Petroleum, mineral and chemical products}). After 2010, industries A (%
\textit{Agriculture and fishing)} and CA (\textit{Food products and
beverages)} join this weakly related group of industries while industry CG (%
\textit{Rubber and plastics}) leaves it.

Table 4 shows the four partition clusters and the main nodes in each
cluster. Despite of the structural evolution of the networks over the 15
years, these are the nodes whose clustering membership remains unchanged.

\begin{center}
\begin{tabular}{|l|l|l|l|l|l|}
\hline
\multicolumn{2}{|l|}{\small Cluster} & {\small Color} & \multicolumn{3}{|l|}%
{\small Main Nodes} \\ \hline
{\small 1} & \textsc{Trade} & {\small green} & {\small G} & {\small A} &
{\small CA} \\ \hline
{\small 2} & \textsc{Construction} & {\small red} & {\small F} & {\small L}
& {\small K} \\ \hline
{\small 3} & \textsc{Manufacture} & {\small blue} & {\small CH} & {\small CK}
& {\small CM} \\ \hline
{\small 4} & \textsc{Arts} & {\small yellow} & {\small R} & {\small JA} &
\\ \hline
\end{tabular}

{\small Table 4: Industrial Clusters.}
\end{center}

The order these clusters are ranked in Table 4 is in line with the results
presented in reference \cite{Blo11} where two measures of node centrality
are defined and applied to data from I/O tables of 37 countries, Portugal
included. The authors showed that, near the year of 2000, \textsc{Trade} is
most frequently the sector with the highest random walk centrality, being
replaced by the \textsc{Construction} cluster in countries like France and
Ireland.

The trees in Figure 3 show that the inter-industry relations suffer a small
change from 2000 to 2005.
\begin{figure}
\begin{center}
\psfig{figure=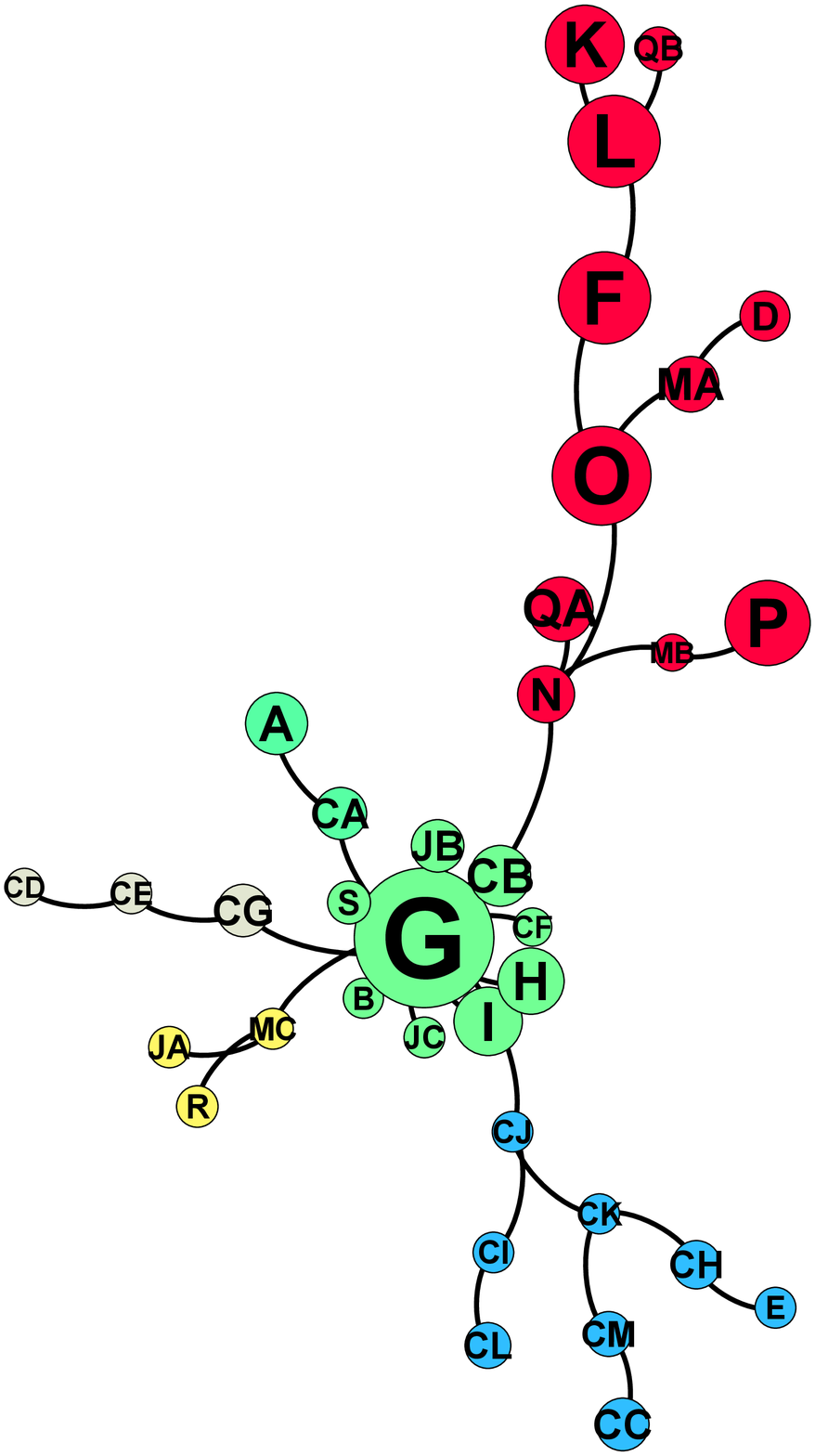,width=6truecm} \psfig{figure=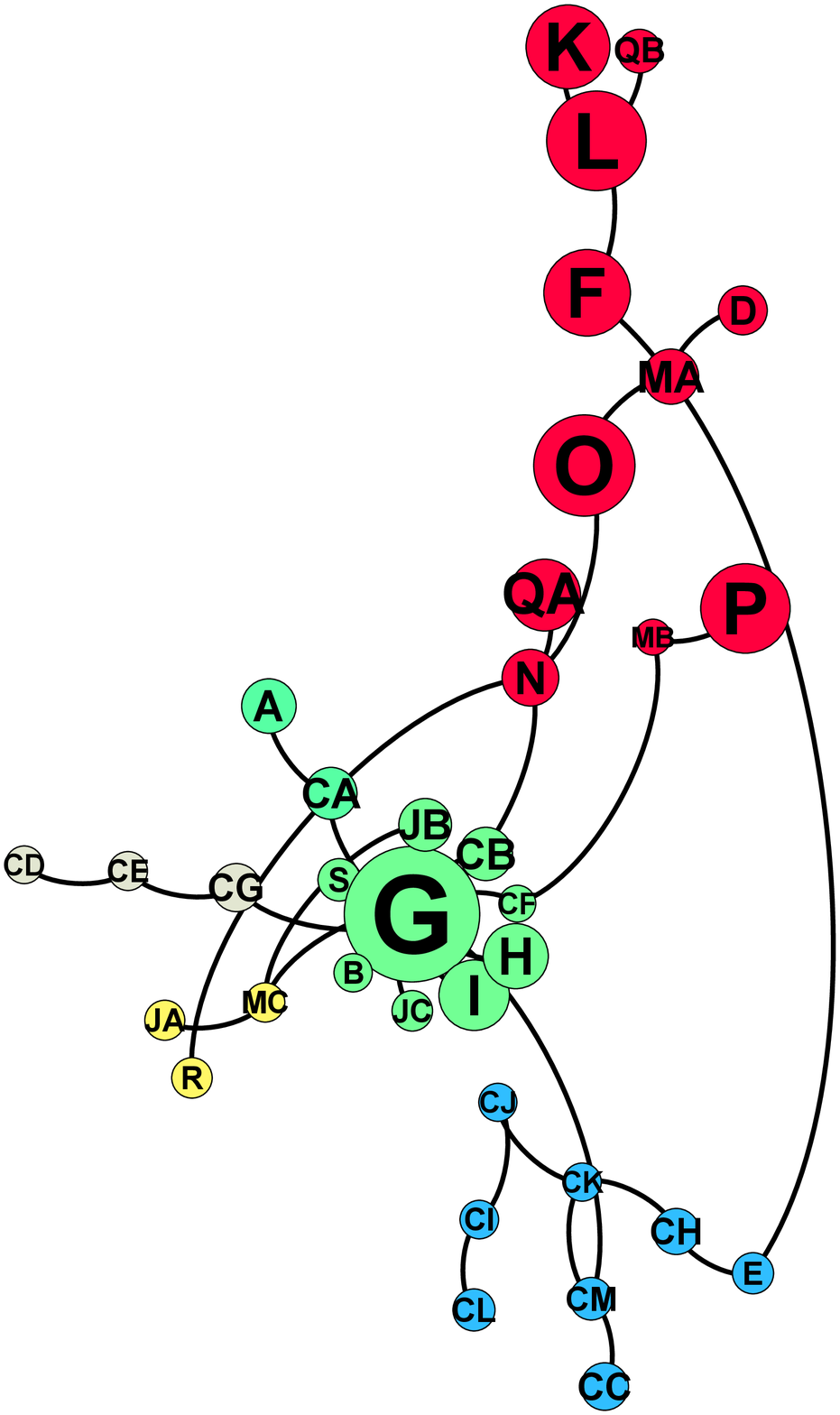,width=6truecm}
\caption{The MST of the networks $N_{36}^{2000}$ and $N_{36}^{2005}$.}
\end{center}
\end{figure}

Being  each of them a MST with 36 nodes, they have exactly 35 links. Let us
define $RL_{2000}^{2005}$ $\in \{0,1,2,...35\}$ as the number of replaced

links from $MST^{2000}$ to $MST^{2005}$. It happens that $RL_{2000}^{2005}$ $%
=5$, due to the few changes occurring from 2000 to 2005, they are: \emph{i)}
industries I (\textit{Accommodation}) and H (\textit{Transportation}) lose
their links to \textsc{Construction} and \textsc{Manufacture} clusters,
respectively, turning both to be connected to \textsc{Trade}; \emph{ii)} in
the opposite direction, industries CI (\textit{Electronics}) and CJ (\textit{%
Electrical equipment}) lose their links to Trade and turn to be connected to
the \textsc{Manufacture} cluster. At last, there is a new link between
industries O (\textit{Public administration}) and N (\textit{Administrative
services}).

In line with the small value of $RL_{2000}^{2005}$, the links inside each
cluster remain almost unchanged from 2000 to 2005. Even the small groups of
white nodes, as the one formed by industries MC (\textit{Other professional}%
, \textit{scientific and technical services}), JA (\textit{Publishing
activities}) and R (\textit{Arts}) and those comprising the core of the
\textsc{Manufacture} cluster (CD, CE and CG) remain without modifications
over that five years period.

Figure 4 shows $MST^{2010}$ and $MST^{2014}${\small . }Looking at the
differences between the networks in the last graph of Figure 3 and the first
graph in Figure 4, we see that from 2000 to 2010, the industries in the
\textsc{Construction} (red) cluster are the most adversely affected being
those with the larger reduction in the number of inter-industry relations.
We envision that is comes from an increase in diversification since these
industries reduced their production values but maintained (or even enlarged)
the diversity of their products.

When the observation of these networks is complemented with the information
on the distribution of the number of products by industry (presented in
Figure 1){\small \ }\ it turns out that the \textit{Construction} industry
seems to be the most affected one. As earlier mentioned, it is a consequence
of the decline in the construction of buildings followed by a corresponding
decrease in \textit{Real estate} (\textit{Leasing and transactions of
buildings}). In that same period, G (\textit{Wholesale trade}) maintains its
production level and reduces the output of the less relevant products.

\begin{figure}
\begin{center}
\psfig{figure=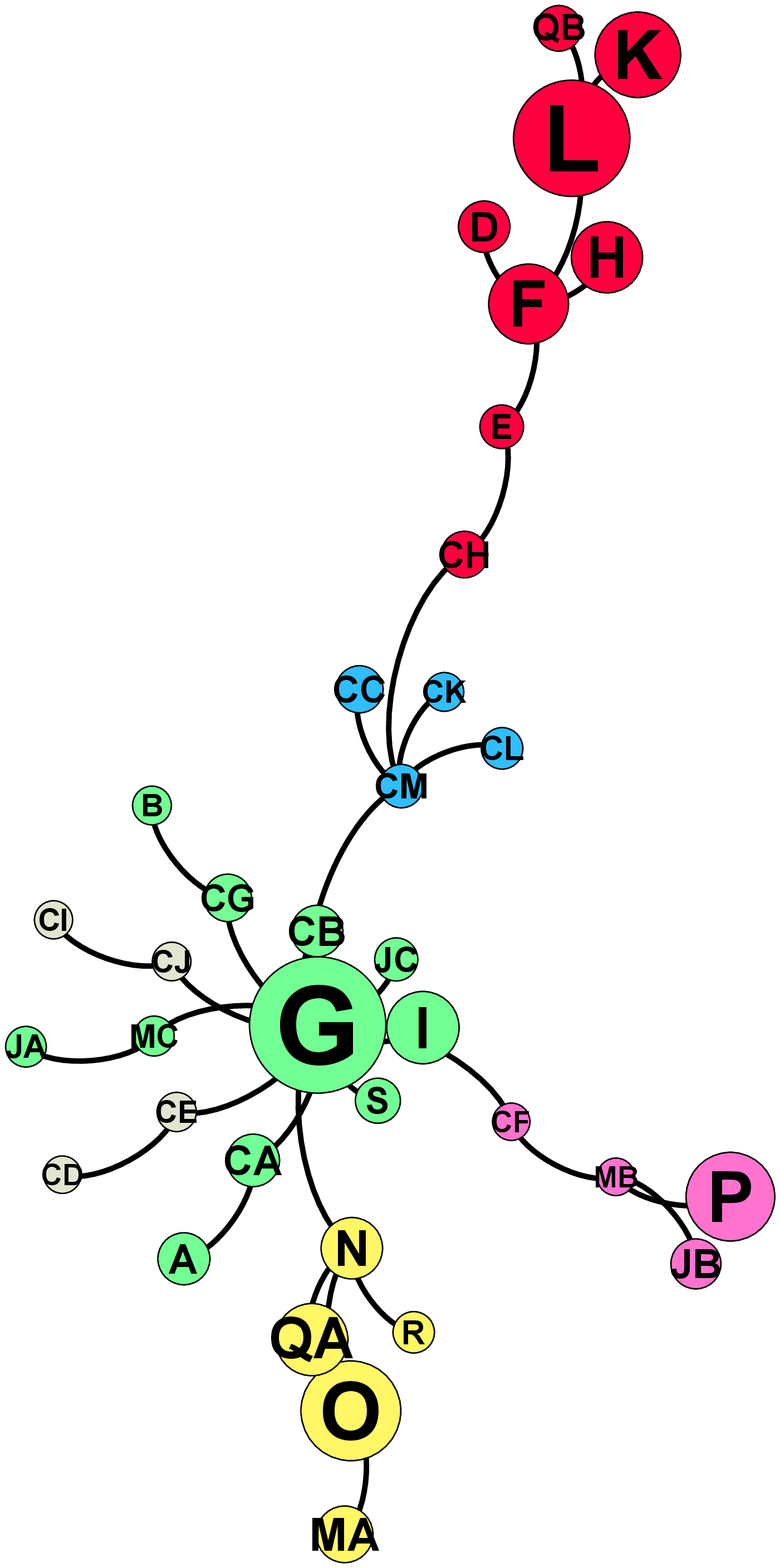,width=6truecm} \psfig{figure=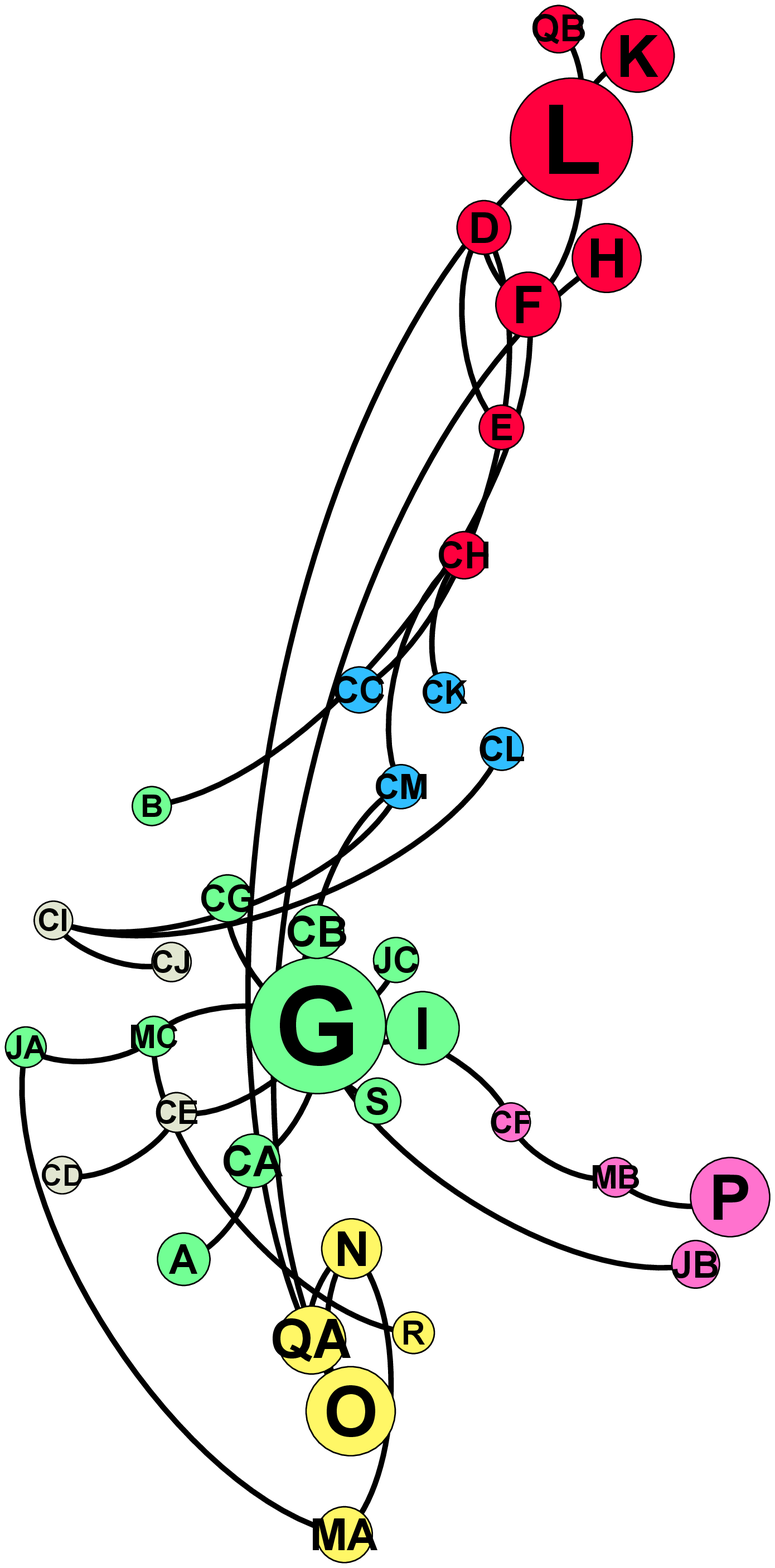,width=6truecm}
\caption{The MST of the networks $N_{36}^{2010}$ and 2014 $N_{36}^{2014}$}
\end{center}
\end{figure}

Conversely to what happened in the comparison of 2000 and 2005, the trees in
Figure 4 show that the inter-industry relations suffer a large change from
2010 to 2014. The number of replaced links reaches 17 ($RL_{2010}^{2014}=17$%
), showing that almost half of the connections is replaced. Such a large
structural modification affects all clusters and brings an entirely new
shape to $MST^{2014}$. The main changes concern: \emph{i)} the breakdown of
the \textsc{Manufacture} (blue) cluster, where almost every link existing in
2010 is removed; \emph{ii) }a great reduction of the number of links inside
the Trade cluster, \emph{iii) }the loss of linkage between the industries MC
(\textit{Other professional, scientific and technical services}) and JA (%
\textit{Publishing}), \emph{iv) }industry CI (\textit{Computer industry})
turns to be linked to the former \textsc{Manufacture} cluster while losing
its links to \textsc{Trade}. Moreover, \emph{v)} the \textsc{Construction}
cluster that shrunk even before 2010 does not recover in 2014 \emph{vi)} the
emergence of a link between the industries MC (\textit{Other professional,
scientific and technical services}) and R (\textit{Arts}) reinforces the
connection between industries in the service sector. Finally, \emph{vii)} in
2014 the link between the industries MB (\textit{R\&D}) and JB (\textit{%
Telecom}) disappears, which can be associated to a decrease in investment in
innovation made by telecom companies. On the other hand, the link between
the industries MB (\textit{R\&D}) and the CF (\textit{Pharmaceutical})
remains, in line with the increasing high level of R\&D participation in the
pharmaceuticals since 2010, when these industries form a new cluster (%
\textit{pink}), as the graphs in Figure 4 show.

To better evaluate the structural changes that characterize the
inter-industry relations, the plots in Figure 5 show the evolution of the
threshold distance ($L^{t}$) in $N_{36}^{t}$\ $t=2001,2002,..,2014$ and the
evolution of the number of replaced links $RL_{t-1}^{t}$ over the same
period.

\begin{figure}[tbh]
\begin{center}
\psfig{figure=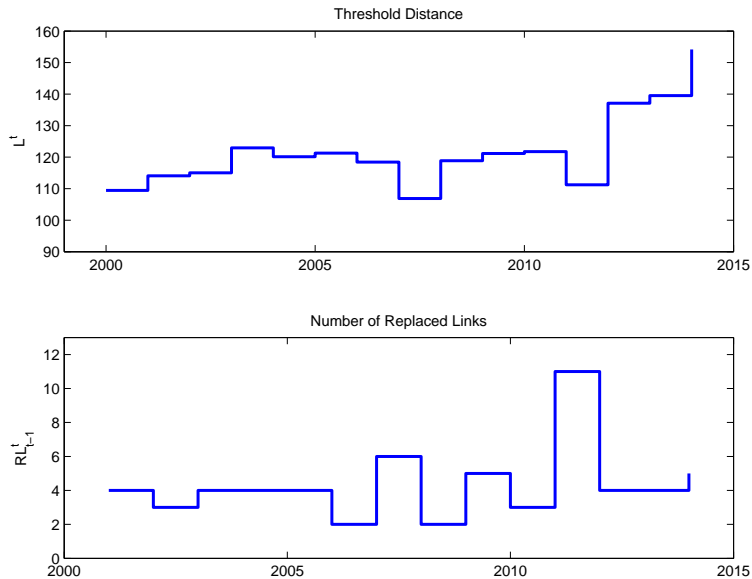,width=10truecm}
\caption{The Threshold Distance $L^{t}$ and the number of  Replaced links $RL_{t-1}^{t}$ in each $MST_{36}^{t}$.}
\end{center}                                                                                                                                                    \end{figure}

The evolution of $L^{t}$ along the 15 years period shows an important
increase of the threshold distance in 2012 after a decrease in 2011. After
2012 the threshold distance value tends to remain almost unchanged and the
corresponding network structures display a small number of replaced links,
as the last years in the plots of Figure 5 show. Increases in the threshold
distance mean that at least one of the network links becomes weaker, leading
to the consideration of a set of larger distances in the hierarchical
clustering process that defines the MST.

Meanwhile, there is a remarkable increase in the number of replaced links
from 2011 to 2012. Earlier we saw that the number of replaced links from
2010 to 2014 ($RL_{2010}^{2014}=17$) reaches 17, showing that almost half of
the connections is replaced in between those four years. Looking to the
evolution of $RL_{t-1}^{t}$ on a yearly-basis suggests that most of the
replacements occurs from 2011 to 2012, over the first year of the Portuguese
economic adjustment.

In fact, it is worthy of attention that those important structural changes
are concentrated in the Portuguese economic adjustment period, providing
therefore enough evidence of the contribute that such network approach
brings to answering our main research question.

\subsection{Beyond the trees}

In the last section, we saw that the large number of links in each network $%
N_{36}^{t}$ makes the extraction of their truly relevant connections a
challenging problem. There, the construction of each MST allowed for the
definition of networks where sparseness replaces high-connectivity in a
suitable way.

However this construction neglects part of the information contained in the
distance matrix, since it only takes the $35$ distances that are considered
in the hierarchical clustering process.

In order to avoid this loss of information we define the projected graph $%
B_{36}^{t}$ (with $N_{36}^{t}$ vertices being the network nodes) by setting $%
b_{i,j}$ $=d_{i,j}$ if $d_{i,j}<$ $L_{36}$ and $b_{i,j}=0$ if $d_{i,j}$ $%
>L_{36}^{t}$. As usual, null arcs of $B_{36}^{t}$ are those for which $%
b_{i,j}=0$. Here we want to consider two nodes $i$ and $j$ to be connected
if $d_{i,j}<$ $L_{36}$.

Let $A_{36}^{t}$ be the boolean graph associated with $B_{36}^{t}$, where
each element $a_{i,j}$ is the number of edges of $B_{36}^{t}$ that join the
vertices $i$ and $j$ and, since $B_{36}^{t}$ is a simple graph, $a$ $\in $$%
\{0,1\}.$

Let us also define $C^{\ast }=\{d_{l}\}$, $l=1,...,m$, as the set of
distances $d_{ij}\in D_{36}^{t}$ whose values are less or equal to $%
L_{36}^{t}$ and compute $S^{t}$

\begin{equation}
S^{t}=m-35
\end{equation}

as the number of \textit{redundant} elements in $C^{\ast }$, that is, the
number of distances $d_{ij}$ that, although being smaller than $L_{36}^{t}$,
\ are not considered in the hierarchical clustering process \cite{Ara00}.
The first plot in Figure 7 shows the evolution of $S^{t}$. Besides the
behavior of $S^{t}$, a detailed way to look at modifications in the number
of redundant elements of the networks can be observed in the following
graphs.

The four plots in Figure 6 show the boolean graphs ($A_{36}^{t}$) obtained
for the years 2000, 2005, 2010 and 2014. They were obtained by:

\begin{enumerate}
\item Taking the matrix of distances ($D_{36}^{t}$) of each period ($%
t=2000,2005,2010,2014$).

\item Applying the hierarchical clustering process to obtain the threshold
distance $L_{36}^{t}$ used in the last step of the hierarchical clustering
process.

\item Building the corresponding boolean graph ($A_{36}^{t}$) where unit
arcs ($d_{ij}^{t}\leq L_{36}^{t}$) are represented as black patches and null
arcs ($d_{ij}^{t}>L_{36}^{t}$) correspond to white ones.
\end{enumerate}

\begin{figure}
\begin{center}
\psfig{figure=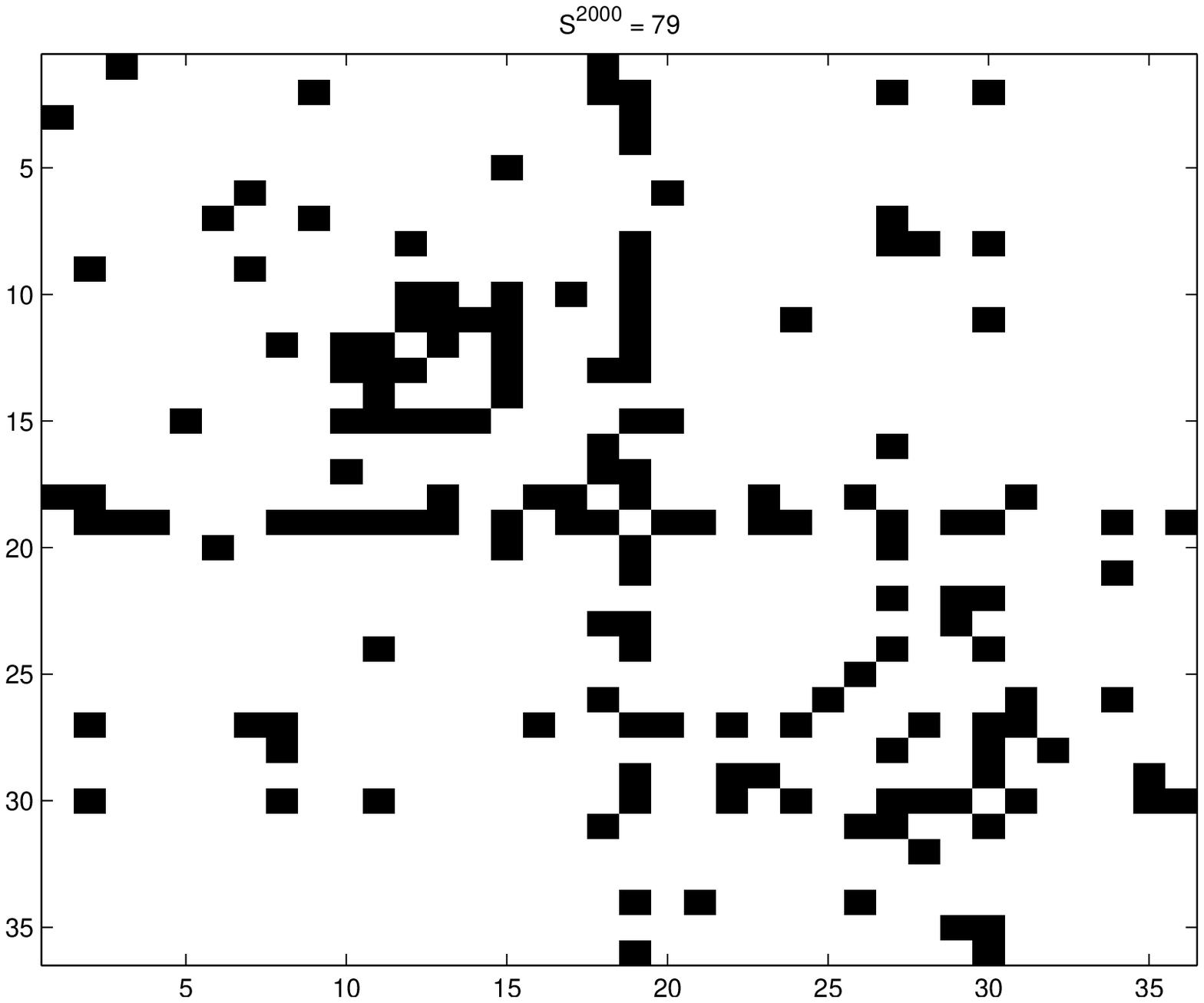,width=5.5truecm}\psfig{figure=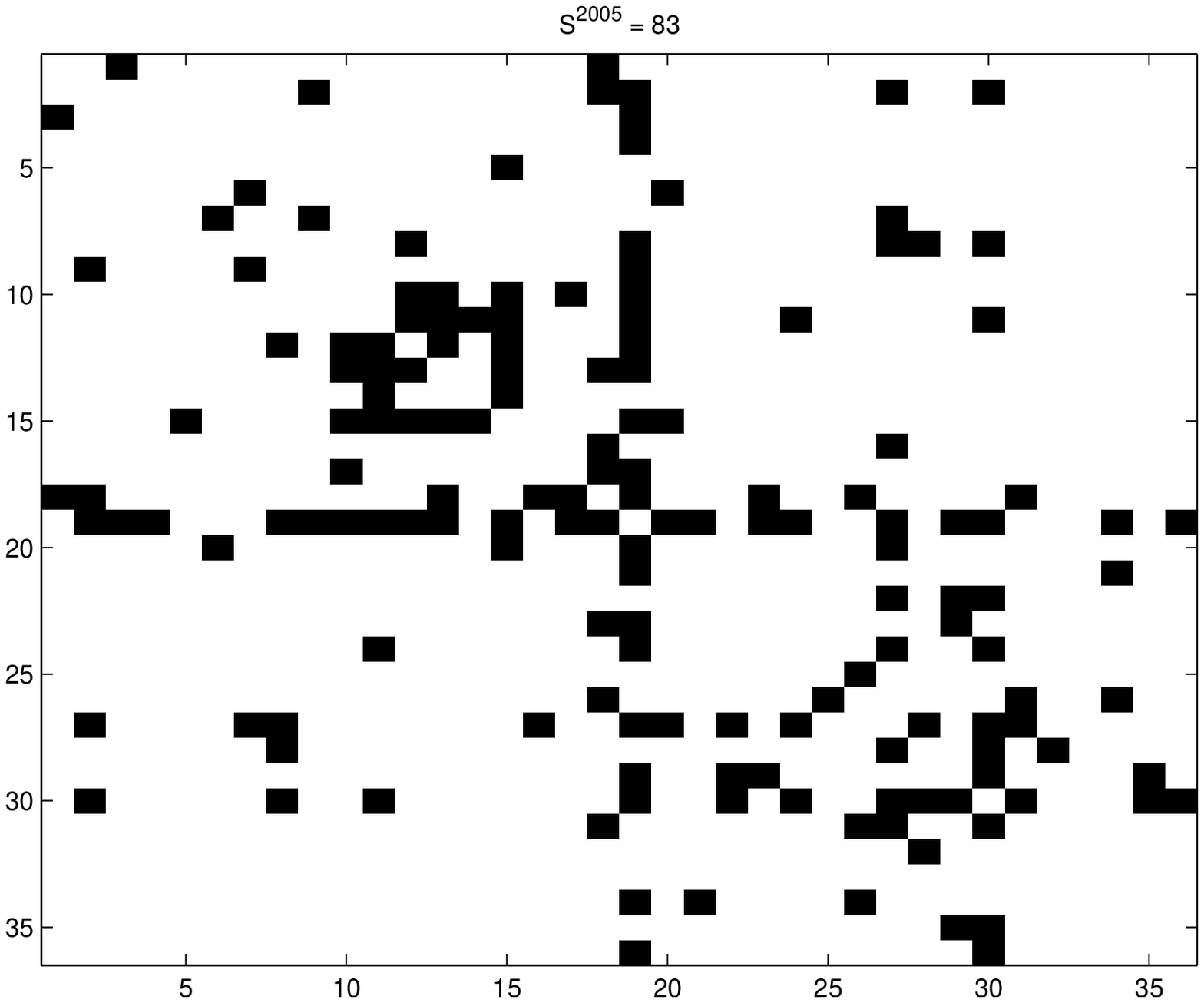,width=5.5truecm}
\psfig{figure=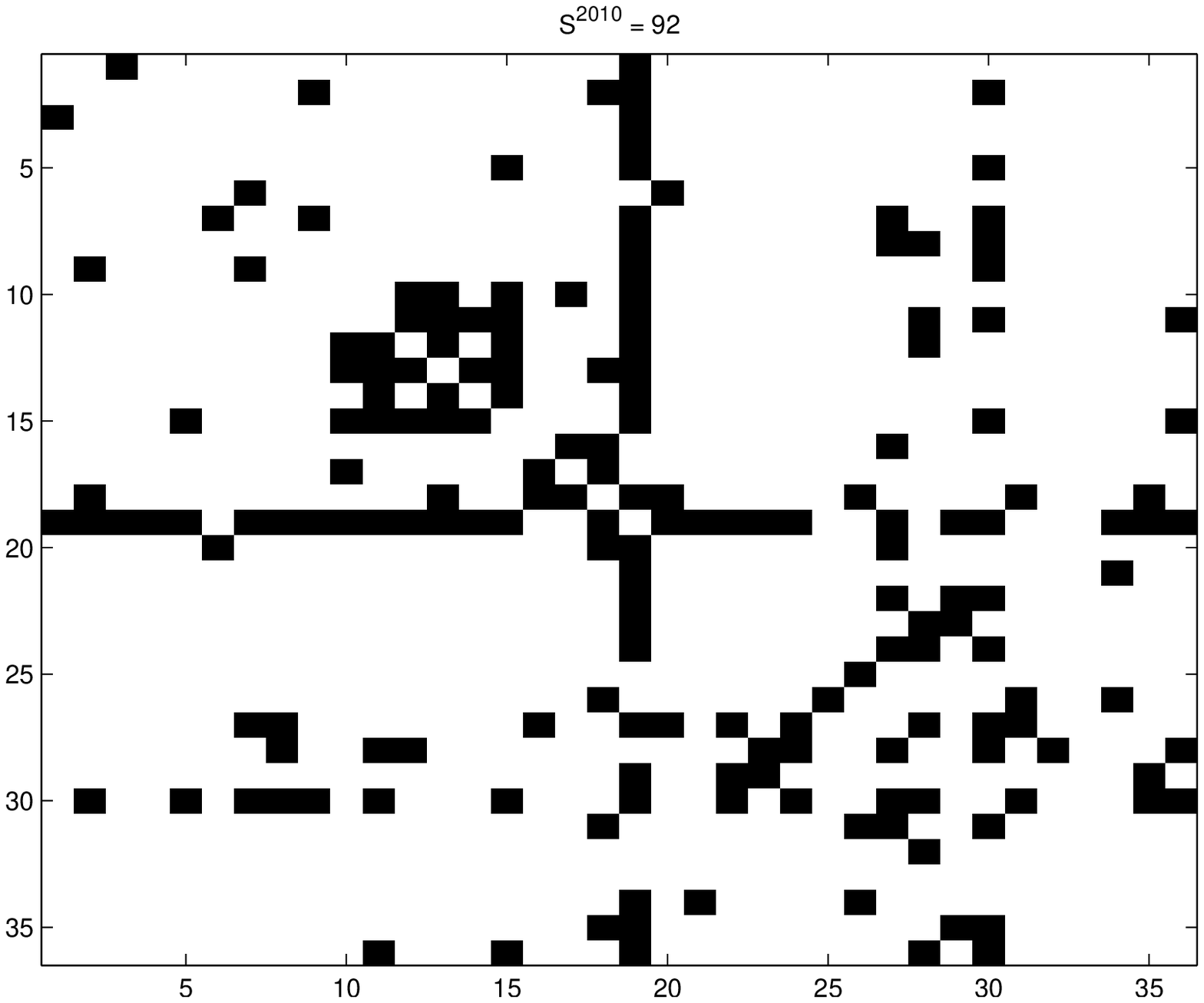,width=5.5truecm}\psfig{figure=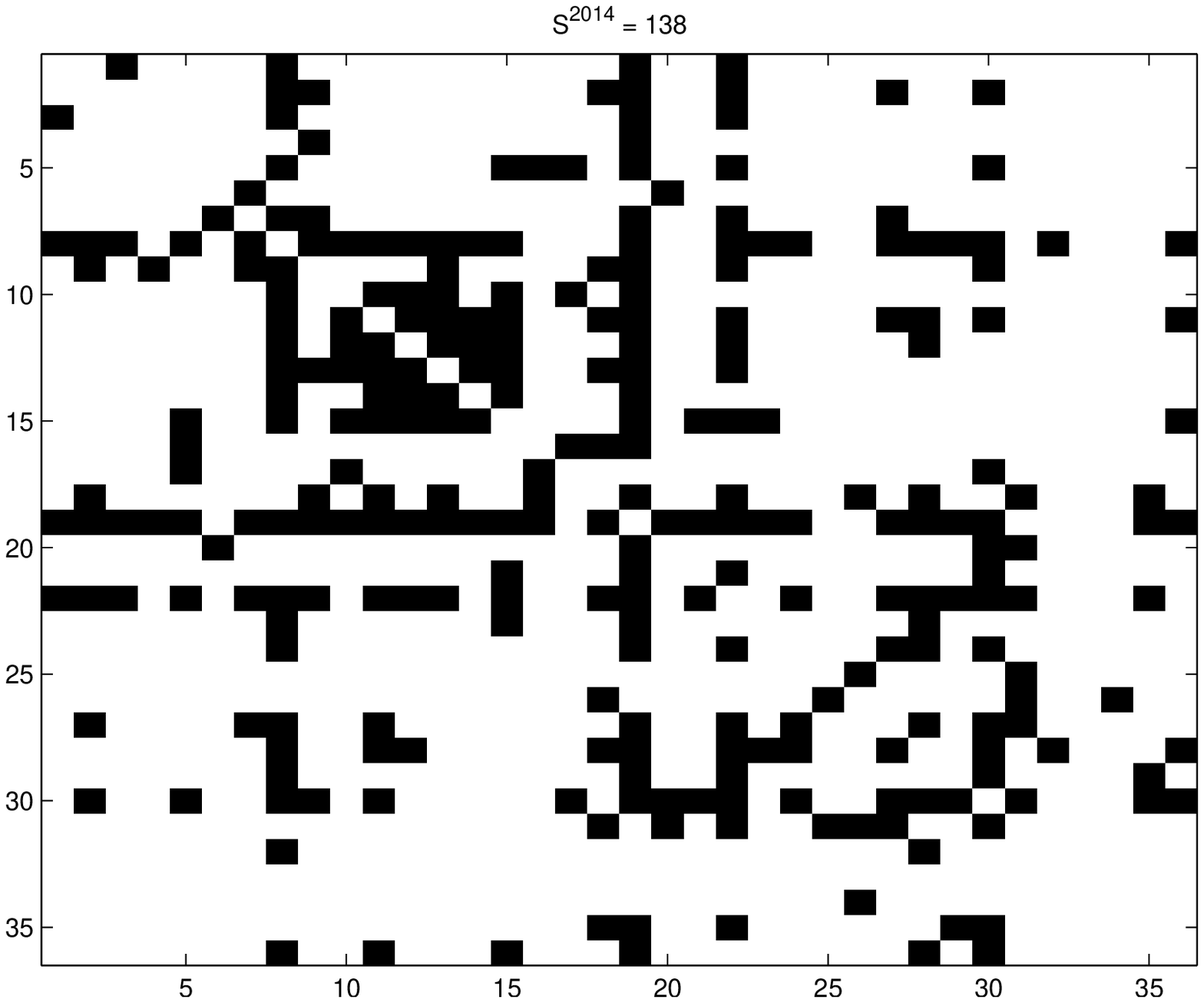,width=5.5truecm}
\caption{The boolean graphs (a) 2000 (b) 2005 (c) 2010 and (d) 2014.}
\end{center}
\end{figure}

The boolean graphs in Figure 6 show the structure of inter-industry
relations in each of the chosen years. There is a relevant difference in
2014, as the graph in Figure 6(d) shows. The number of black patches
strongly increases in the structure represented in the last graph of Figure
6, showing that the number of unit arcs is much greater than 35.
Consequently, the number of redundant elements in the graph is significantly
greater than in the other three graphs. As shown in this figure, the network
structure in 2014 tend to contain cycles and moves away from the tree-like
structures that characterize those obtained for 2000, 2005 and 2010 (graphs
(a), (b) and (c) in Figure 6).

As we aim at characterizing the richer connectivity structure that emerges
in 2014, besides the topological coefficient $S^{t}$ that captures the
number of redundant elements in the networks, we compute the residuality
coefficient $R^{t}$ \cite{Ara00}, targeted at measuring the amount of
residuality in each network $N_{36}^{t}$.

\begin{equation}
R^{t}=\frac{\sum_{d_{i,j}>L}d_{i,j}^{-1}}{\sum_{d_{i,j}\leq L}d_{i,j}^{-1}}
\end{equation}

where $L_{36}^{t}$ is the threshold distance value that insures connectivity
of the whole network in the hierarchical clustering process.

The evolution of $S^{t}$ along the 15 years period is presented in the first
plot of Figure 7. The value of $S^{2014}$=138 shows an important increase in
the number of redundant links. Our previous studies (\cite{Ara07},\cite%
{Ara16b},\cite{Ara00},\cite{Vil16}) revealed that the observation of an
important increase in redundancy use to be a consequence of disturbed
periods being the topological correlate of the occurrence of stress events.
Here, the large increase in the number of redundant elements from 2012 to
2014 provides enough evidence of the structural changes taking place on the
network structure after the implementation of the economic adjustment
program in Portugal.

The coefficient of residuality relates the relative strengths of the links
above and below the threshold distance value ($L_{36}^{t}$). In other words,
it relates the relative strengths of weak and strong links. Increases in $%
R_{36}^{t}$ usually happen because strong ties become even stronger (short
distances are shortened). Moreover, it also comes from the fact that\ the
networks become less sparse (the number of links increases), forcing several
weak links to leave this category. These structural reinforcement is often
associated with the occurrence of stress events, as the implementation of
the recent process of Portuguese economic adjustment.

\begin{figure}
\begin{center}
\psfig{figure=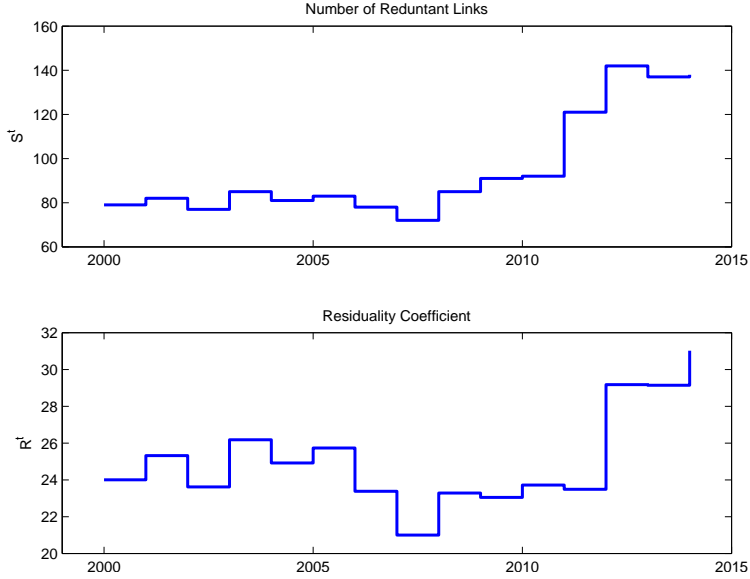,width=10truecm}
\caption{The number of Redundant links $S^{t}$ in the networks $N_{36}^{t}$ and the Residuality Coefficient $R^{t}$.}
\end{center}
\end{figure}

The plots in Figure 7 show the evolution of the residuality coefficient\ ($%
R^{t}$) in the networks $N_{36}^{t}$\ $t=2000,2001,..,2014$ and the
evolution of the ($S^{t}$) in the same time interval. They allow for
capturing the main differences in the behavior of these coefficients in the
latest years. Indeed, the plots in Figure 7 show that the lowest values of $%
S^{t}$ and $R^{t}$ correspond to the years before 2011. After 2011, as
revealed by the strong increase of $S^{t}$ and $R^{t},$ a major change is
occurring in the structure of the networks $N_{36}^{2012},N_{36}^{2013}$ and
$N_{36}^{2014}$. This change, being empirically described, suggests that in
the recent process of economic adjustment suffered by the Portuguese
economy, a new structure emerges from the inter-industry relations.

\section{Conclusion}

The literature has shown an increasing use of Input/Output tables to induce
networks of money flows between industries in national economies. Because
I/O tables are quite similar to adjacency matrices, their representation as
networks intuitively emerges. Here we follow a similar approach but instead
of the usual choice of I/O tables, our data comes from Industry/Product
Output tables and the induced networks are proximity networks. Consequently,
the industries are not explicitly linked by any concrete relation existing
in the real world but for a well-defined measure of similarity. Although the
induction of proximity networks is less intuitive than those obtained from
I/O tables, they provide useful analytical settings, making it possible to
investigate the structural properties of economic networks in a suitable way.

Furthermore, as the networks induced from the output tables compiled by the
Portuguese national accounting agency are year-based, they provide a
convenient framework to evaluate the impact of the sovereign debt crisis and
the implementation of the economic and financial adjustment program in
Portugal (2011-2014).

Our conclusions can be summarized in the following.

\begin{enumerate}
\item the \textit{Wholesale Trade} (G) industry is the most central and also
the most connected node in the network of 36 industries. The nodes connected
to G form the biggest and the most stable cluster in the entire 15 years
period. Among the few modifications that took place in G in the last years
is the weakening of its connection to the \textit{Pharmaceutical }(CF)
industry. The weakening of the link between Trade and the Pharmaceuticals
can be attributed to the reduction of profit margins in the prices
negotiated by the Ministry of Health with the pharmaceutical industry. Along
the financial adjustment program, such a negotiation envisioned a
substantial reduction of the expense with hospital drugs.

\item the \textit{Construction} (F) industry suffer an important shrunk due
to the decline in the construction of buildings in Portugal, being followed
by a even greater decrease in the \textit{Real estate} (L) industry.

\item in 2014 the link between the industries \textit{R\&D} (MB) and \textit{%
Telecom} (JB) disappears, which is in line with an important decrease of
investment in innovation made by telecom companies.

\item on the other hand, from 2010 to 2014, the link between the industries
\textit{R\&D} (MB) and the \textit{Pharmaceuticals} (CF) remains, converging
with the increasing and high level of R\&D participation in the
pharmaceuticals since 2010.

\item the emergence of a link between the industries \textit{Scientific and
technical services} (MC) and \textit{Arts} (R)\ in 2014 shows the
reinforcement of an important relation holding industries in the service
sector (to the detriment of those in the manufacturing and agricultural
ones).
\end{enumerate}

The richer connectivity structure that emerges in 2014 was characterized by
two topological coefficients: the one that captures the number of redundant
elements in the networks, and that measuring their amount of residuality.
The investigation on the temporal evolution of these coefficients showed
that:

\begin{enumerate}
\item[a) ] important structural changes took place on the industrial
networks along the implementation of the economic and financial adjustment
program in Portugal, as indicated by the large increase in the number of
redundant elements in the networks from 2012 to 2014. It is worth noticing
that the observation of an important increase in redundancy use to be a
consequence of disturbed periods being the topological correlate of the
occurrence of stress events.

\item[b)] the important increase in the value of the coefficient of
residuality shows that in 2011, the relative strengths of the links above
and below the threshold distance value is higher than in any other year. As
it relates the relative strengths of weak and strong links, increases in $%
R_{36}^{t}$ are usually due to the shortening of distances, forcing several
weak links in the networks to leave this category. These structural
reinforcement is often associated with the occurrence of stress events, as
the recent process of Portuguese economic adjustment.
\end{enumerate}

The results obtained here confirm that the network analysis does indeed
provide relevant information on the structural changes that characterize the
three years of economic adjustment in Portugal. These results contribute to
illustrate the usefulness of inducting similarity networks from output
tables and the consequent promising power of the graph formulation for the
analysis of inter-industry relations.

\pagebreak

\newpage
\subsection*{Appendix  {\small Table A: Industries and products according to NACE and CPA codes (I38\&P38).}}
\begin{tabular}{|l|c|c|c|}
\hline
{\small NACE/CPA codes } & {\small Description} & {\small I38} & {\small P38}
\\ \hline
\multicolumn{1}{|c|}{\small 01-03} & {\small Agriculture, forestry and
fishing } & {\small A} & {\small A} \\ \hline
\multicolumn{1}{|c|}{\small 05-09} & {\small Mining and quarrying} & {\small %
B} & {\small B} \\ \hline
\multicolumn{1}{|c|}{\small 10-12} & {\small Food products, beverages and
tobacco products} & {\small CA} & {\small CA} \\ \hline
\multicolumn{1}{|c|}{\small 13-15} & {\small Textiles, wearing apparel,
leather and related products} & {\small CB} & {\small CB} \\ \hline
\multicolumn{1}{|c|}{\small 16-18} & {\small Wood and paper products, and
printing} & {\small CC} & {\small CC} \\ \hline
\multicolumn{1}{|c|}{\small 19} & {\small Coke and refined petroleum products%
} & {\small CD} & {\small CD} \\ \hline
\multicolumn{1}{|c|}{\small 20} & {\small Chemicals and chemical products} &
{\small CE} & {\small CE} \\ \hline
\multicolumn{1}{|c|}{\small 21} & {\small Basic pharmaceutical products and
pharmaceutical preparations} & {\small CF} & {\small CF} \\ \hline
\multicolumn{1}{|c|}{\small 22-23} & {\small Rubber and plastic products} &
{\small CG} & {\small CG} \\ \hline
\multicolumn{1}{|c|}{\small 24-25} & {\small Basic metals and fabricated
metal products, except machinery and equipment} & {\small CH} & {\small CH}
\\ \hline
\multicolumn{1}{|c|}{\small 26} & {\small Computer, electronic and optical
products} & {\small CI} & {\small CI} \\ \hline
\multicolumn{1}{|c|}{\small 27} & {\small Electrical equipment} & {\small CJ}
& {\small CJ} \\ \hline
\multicolumn{1}{|c|}{\small 28} & {\small Machinery and equipment n.e.c.} &
{\small CK} & {\small CK} \\ \hline
\multicolumn{1}{|c|}{\small 29-30} & {\small Motor vehicles, trailers and
semi-trailers} & {\small CL} & {\small CL} \\ \hline
\multicolumn{1}{|c|}{\small 31-33} & {\small Furniture and other
manufacturing} & {\small CM} & {\small CM} \\ \hline
\multicolumn{1}{|c|}{\small 35} & {\small Electricity, gas, steam and air
conditioning} & {\small D} & {\small D} \\ \hline
\multicolumn{1}{|c|}{\small 36-39} & {\small Water collection, treatment and
supply} & {\small E} & {\small E} \\ \hline
\multicolumn{1}{|c|}{\small 41-43} & {\small Construction} & {\small F} &
{\small F} \\ \hline
\multicolumn{1}{|c|}{\small 45-47} & {\small Wholesale and retail trade } &
{\small G} & {\small G} \\ \hline
\multicolumn{1}{|c|}{\small 49-53} & {\small Transportation} & {\small H} &
{\small H} \\ \hline
\multicolumn{1}{|c|}{\small 55-56} & {\small Accommodation, food and
beverage service} & {\small I} & {\small I} \\ \hline
\multicolumn{1}{|c|}{\small 58-60} & {\small Publishing, audiovisual and
broadcasting} & {\small JA} & {\small JA} \\ \hline
\multicolumn{1}{|c|}{\small 61} & {\small Telecommunications} & {\small JB}
& {\small JB} \\ \hline
\multicolumn{1}{|c|}{\small 62-63} & {\small Computer programming and
consultancy} & {\small JC} & {\small JC} \\ \hline
\multicolumn{1}{|c|}{\small 64-66} & {\small Financial and Insurance} &
{\small K} & {\small K} \\ \hline
\multicolumn{1}{|c|}{\small 68} & {\small Real estate} & {\small L} &
{\small L} \\ \hline
\multicolumn{1}{|c|}{\small 69-71} & {\small Legal, accounting. management
consulting, architectural and engineering services} & {\small MA} & {\small %
MA} \\ \hline
\multicolumn{1}{|c|}{\small 72} & {\small Scientific research and development%
} & {\small MB} & {\small MB} \\ \hline
\multicolumn{1}{|c|}{\small 73-75} & {\small Other professional, scientific
and technical services} & {\small MC} & {\small MC} \\ \hline
\multicolumn{1}{|c|}{\small 77-82} & {\small Administrative and support
services} & {\small N} & {\small N} \\ \hline
\multicolumn{1}{|c|}{\small 84} & {\small Public administration and defense}
& {\small O} & {\small O} \\ \hline
\multicolumn{1}{|c|}{\small 85} & {\small Education} & {\small P} & {\small P%
} \\ \hline
\multicolumn{1}{|c|}{\small 86} & {\small Human health} & {\small QA} &
{\small QA} \\ \hline
\multicolumn{1}{|c|}{\small 87-88} & {\small Social work} & {\small QB} &
{\small QB} \\ \hline
\multicolumn{1}{|c|}{\small 90-93} & {\small Arts, entertainment and
recreation} & {\small R} & {\small R} \\ \hline
\multicolumn{1}{|c|}{\small 95} & {\small Other services } & {\small S} &
{\small S} \\ \hline
\multicolumn{1}{|c|}{\small 97-98} & {\small Households as employers of
domestic personnel} & {\small T} & {\small T} \\ \hline
\multicolumn{1}{|c|}{\small 99} & {\small Activities of extraterritorial
organizations and bodies} & {\small U} & {\small U} \\ \hline
\end{tabular}
\end{document}